# DotDFS: A Grid-based high-throughput file transfer system

Alireza Poshtkohi[a], M.B. Ghaznavi-Ghoushchi [a,*]

[a]Department of Electrical Engineering, Shahed University, Tehran 331911865, Iran

**Abstract**

DotGrid platform is a Grid infrastructure integrated with a set of open and standard protocols recently implemented on the top of Microsoft .NET in Windows and MONO .NET in UNIX/Linux. DotGrid infrastructure along with its proposed protocols provides a right and solid approach to targeting other platforms, e.g., the native C/C++ runtime. In this paper, we propose a new file transfer protocol called DotDFS as a high-throughput distributed file transfer component for DotGrid. DotDFS introduces some open binary protocols for efficient file transfers on current Grid infrastructures. DotDFS protocol also provides mechanisms for multiple file streams to gain high-throughput file transfer similar to GridFTP protocol, but by proposing and implementing a new parallel TCP connection-oriented paradigm. In our LAN tests, we have achieved better results than Globus GridFTP implementation particularly in multiple TCP streams and directory tree transfers. Our LAN experiences in memory-to-memory tests show that DotDFS accesses to the 94% bottleneck bandwidth while GridFTP is accessing 91%. In LAN disk-to-disk tests, comparing DotDFS protocol with GridFTP protocol unveils a set of interesting and technical problems in GridFTP for both the nature of the protocol and its implementation by Globus. In the WAN experimental studies, we propose a new idea for analytical modeling of file transfer protocols like DotDFS inspired by sampling, experimentation and mathematical interpolation approaches. The cross-platform and open standard-based features of DotDFS provide a substantial framework for unifying data access and resource sharing in real heterogeneous Grid environments.

## 1. Introduction

Recently, integrated use of system resources are utilized by Grid and Cloud platforms. Grid [1,2,3] infrastructures provide the ability to share, select and aggregate distributed resources as computers, storage systems or other devices in an integrated way. Effective end-to-end transmission of data demands a system approach in which file systems, computers, network interfaces, and network protocols are managed in an integrated fashion to meet performance and robustness goals [4].

Secure and high-speed data transfers are vital and profound in Grid environments. GridFTP protocol is recently used as the de-facto standard for bulk data transfers in various Grid projects in high-bandwidth wide-area networks all over the world [4,5,6]. Globus [4] has developed GridFTP protocol for UNIX-class style operating systems [4] , and a team at the University of Virginia implemented GridFTP for Windows-based operating systems via Microsoft .NET Platform [7]. Globus has fully implemented the GridFTP protocol with the following major features: third-party control of data transfer, authentication, data integrity, data confidentiality, striped data transfer, parallel data transfer, partial file transfer, automatic negotiation of TCP buffer/window sizes, and support for reliable and restartable data transfer.

---

* Corresponding author.
 *E-mail addresses:* alireza.poshtkohi@gmail.com (A. Poshtkohi), ghaznavi@shahed.ac.ir (M.B. Ghaznavi-Ghoushchi).

[4,5,6] describe GridFTP protocol and its full features. The authors in [7] also addressed the implementation problems of GridFTP for .NET and Windows platforms. Some GridFTP problems formerly studied by researchers are long-time GridFTP setup due to any application setup, connection and authentication, firewalls, and network address translation [8,9,10,11].

We have developed DotGrid platform [12,13,14] which enables the creation of Grid infrastructure using Microsoft .NET [15] in Windows and MONO .NET [16] in UNIX and Linux environments. It provides Grid services and toolkits for fast developing Grid applications.

The main goal of DotGrid project [12,13,14] is to develop a cross-platform framework for sharing computational resources between heterogeneous DotGrid nodes. To build a cross-platform desktop Grid infrastructure in heterogeneous platforms, DotGrid makes use of high-performance implementation of the native .NET Framework in Windows and MONO .NET project in UNIX/Linux family of operating systems, respectively. Resource sharing and bulk data transfer are two major architectural needs that we have investigated during the design process of DotGrid.

Currently, we are implementing a new cross-platform Grid/Cloud infrastructure, which extends and utilizes the open and standard protocols suggested in DotGrid platform for native runtime, i.e., C/C++ stack. This new platform is applying the ECMA international open standards [33,34] to provide a much more standardized native implementation of DotGrid for scientific and enterprise Grid/Cloud communities in the most recently operating systems including UNIX, Linux, and Windows.

Recently, WAN-based file systems like Lustre [18], GPFS-WAN [19] and Gfarm [20] are maturated. Due to our knowledge, it seems that DotDFS is more like to GridFTP than the above mentioned file systems. For example, DotDFS may be used, as well as GridFTP, in Gfarm as the underlying file transfer protocol. DotDFS has many different aspects ranging from structure to architecture with GridFTP protocol. Most of them are declared during this paper.

In this paper, we introduce a Grid-based high-throughput file transfer system, called DotDFS. This can be used as a component of a computational Grid [12,13,14] or data Grid [17] environments.

The rest of the paper is organized as follows. In section two, we present DotDFS protocol. Section 3 infers high-performance server design patterns for the DotDFS protocol. Sections four and six describe DotDFS implementation and DotSec GSI. Section five focuses on comparison of DotDFS protocol with GridFTP protocol. The LAN and WAN experimental studies are described in section 7. We conclude in section 8.

## 2. DotDFS Protocol

The current version of DotDFS protocol is a binary protocol model like TCP/IP, DNS and SOCKS protocols. This platform-independent feature allows DotDFS protocol to be re-implemented in other platforms and gains more interoperability. This approach results in more performance and throughput in the cost of implementation complexity.

[7] reports interoperability problems of the Globus GridFTP server due to the MPI used in Globus core. The two major problems in the current .NET GridFTP implementation are: no support of the authentication in data channels and no interoperability of the stripped data transfers with the Globus GridFTP. The .NET GridFTP extensively used from Windows services and native Win32 APIs. This causes it could not be ported to UNIX/Linux platforms via MONO .NET.

DotDFS is a high-throughput file transfer protocol which meets the requirements to set it as a background for applications in the areas of distributed, cluster, grid and cloud computing. These requirements are discussed in this section and have been considered in our current DotDFS implementation.

DotDFS heavily depends upon a set of abstractions that are used to hide infrastructure dependencies. DotDFS APIs can be used for implementing applications in high-throughput data transfers similar to GridFTP. Other applications that can be implemented relied upon on DotDFS are: data intensive Grid applications, Grid-based resource sharing, booting operating systems from high-speed networks, peer-to-peer file systems and distributed databases.

In DotDFS protocol, upon establishment of a connection to a DotDFS server, each client negotiates its favorite session with a set of flexible parameters. The topological architecture of DotDFS protocol is depicted in Fig. 1. The following paragraphs present the details of the protocol.

The default port of DotDFS server is 2799. When a client starts to negotiate with the server, it sends a one-byte binary header to notify the server from its requested service type. In the current version of DotGrid, values of 0, 1, and 2 are defined for specifying the request modes of DotDfsMode, DotGridRemoteProcessMode and DotGridThreadMode service, respectively.

DotDFS protocol requires that all established connections to the server must be authenticated and authorized after service selection through DotSec GSI and DotSec TSI. This follows with X-Channels or TSI X-Channels selection between client and server. DotSec protocol is explained in section 6 with more details. The current version of DotDFS supports three servicing modes from the viewpoint of the connected clients to a DotDFS server including DFSM, FTSM, and PathM.

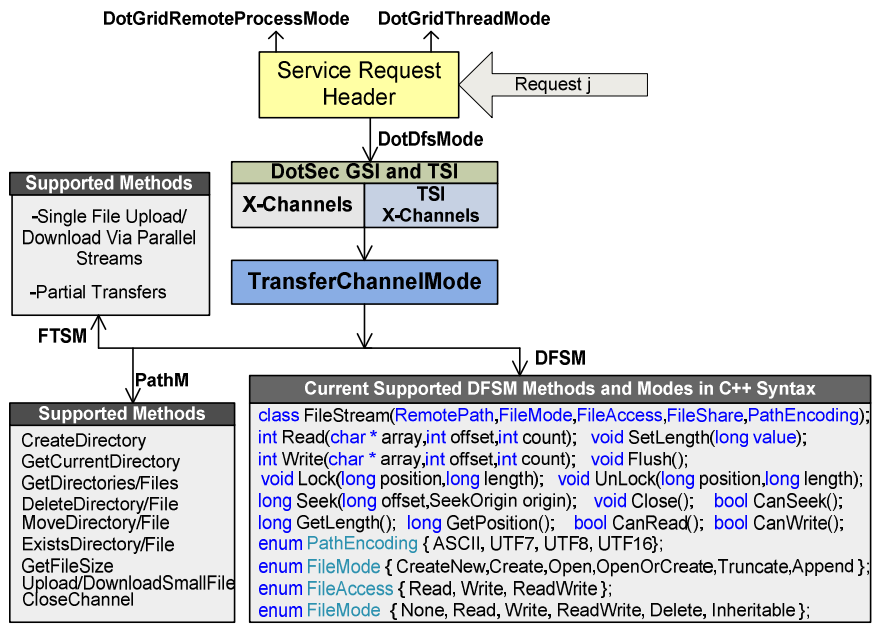

**Fig. 1.** Topological DotDFS protocol representation.

## 2.1. Distributed File System Mode (DFSM)

This requested mode supports the access to the files and data sharing mechanisms which are used in conventional distributed file systems. Moreover, this mode can be used for stripped and third-party data transfers. One good example of this mode is a situation with one or more transport streams between *m* network endpoints sending side and *n* network endpoints on the receiving side. In [12,13], we developed a distributed cryptographic engine to encrypt/decrypt and transfer petabytes scale files by using data striping approaches.

The major of DFSM origination is inspired by the unique features of FileStream class and System.IO namespace declarations for file/directory manipulations stated in Microsoft .NET Framework 1 and later versions. FileStream features in file IO streaming is the main reason for selecting it for DFSM mode of DotDFS protocol.

Due to the wide range of methods and arguments shown in Fig. 1 (like *Seek()*, *Read()*, and *Flush()*), DotDFS protocol not only supports POSIX file semantics, but also it can be used in other scope of applications like remote file streaming and remote named pipes for inter-process communications (IPC). Recently, we are investigating some approaches to fully supporting this feature relied upon DotDFS protocol for unifying IPC's via in UNIX/Linux and Windows operating systems on Grids. Access to NFS, Microsoft DFS (this will be discussed with PathM mode in section 2.3) and other pluggable file systems are also the applications of DFSM mode.

Here is an example for the binary model feature of the DotDFS protocol. Let's have a DotDFS session with a Windows platform client and a Linux platform server. The DFSM mode for copying a 100MB remote file to local storage is examined. The C++ pseudo code of this scenario is shown in Fig. 2.

The required binary header for *remote->Read()* method of Fig. 2 is illustrated in Fig. 3. DotDFS server replies to the *Read()* request of client with the value of zero in Method block. This value is used for declaring the response of *Read()* request. The length of transmitted buffer is stored in RW-Length field. RW-Mode specifies the value of RW-Length in turn. We have applied a different approach in this header and in spite of all other binary protocols with RW-Length default value of 4-byes, we set it variable and controlled with the actual length of the read buffer from the storage system.

For example, if the submitted buffer size is less than $2^{16}$ and greater than $2^8$, then RW-Mode is filled with value of 2. This technique in buffer space strength reduction is widely used during the design of our proposed DotDFS protocol.

```
int read = 0;
char *buffer = new char[256*1024];
NetworkCredential *nc = new NetworkCredential("user", "pass");
DotDfsFileStream *remote = new DotDfsFileStream("/root/files/100meg.dat",
        FileMode.Open, FileAccess.Read, FileShare.None,
        PathEncoding.UTF8, "192.168.1.2", nc, false);
        // remote file stream creation via DFSM mode
FileStream *local = new FileStream("D:\\100meg.dat", FileMode.Create,
        FileAccess.Write, FileShare.None); // local file stream creation
while((read = remote->Read(buffer, 0, buffer.Length)) != 0)
{
    local->Write(buffer, 0, read); //copies remote read file blocks
}
remote->Close();
local->Close();
```

**Fig. 2.** DFSM Mode pseudo code sample.

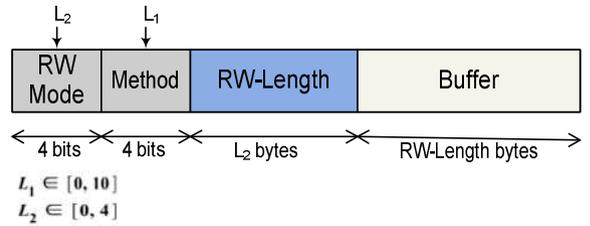

**Fig. 3.** Read/Write method header for DFSM mode.

## 2.2. File Transfer System Mode (FTSM)

In this mode, DotDFS protocol not only supports the Grid computing demands investigated by GridFTP developer teams [4,5,6] but also includes procedures and methods for high-throughput and high-performance in dedicated client-server architectures based on DotDFS protocol. FTSM mode results in high throughput and performance file transfer in DotDFS protocol. The following paragraphs describe the major features of FTSM.

DotDFS protocol defines a set of specifications for increasing file transfer throughput through establishing multiple TCP connections in parallel to accelerate start-up in the TCP slow rate, and negotiating the TCP Socket Window buffer size between a DotDFS server and client before starting the DotDFS session according to the bandwidth-delay product of a network. This case meets when X-Channels are changed to be TCP/IP channels as shown in Fig. 1. It must be noticed that in DotDFS protocol, eXtensible channels are an abstraction concept for low-level transmission control protocols between endpoints. The TCP/IP stack is the main default protocol used in DotGridSocket layer in which all network I/O requests are processed in the specifications and implementation of the DotDFS protocol. Due to the WAN-based TCP overheads, the SCTP protocol [32] will be considered as a replacement for TCP protocol in our future research development.

Fig. 4 shows the client-server DotDFS protocol communications in FTSM mode. All the shown communications are binary forms. Initially client sends a FTSM service selection header to DotDFS server and requests the service type from DotDFS server. Then, if there is an available service, the server sends an available DotDFS service header, and puts the client in authentication stage and negotiation enforcement for sending DotSec TSI parameters.

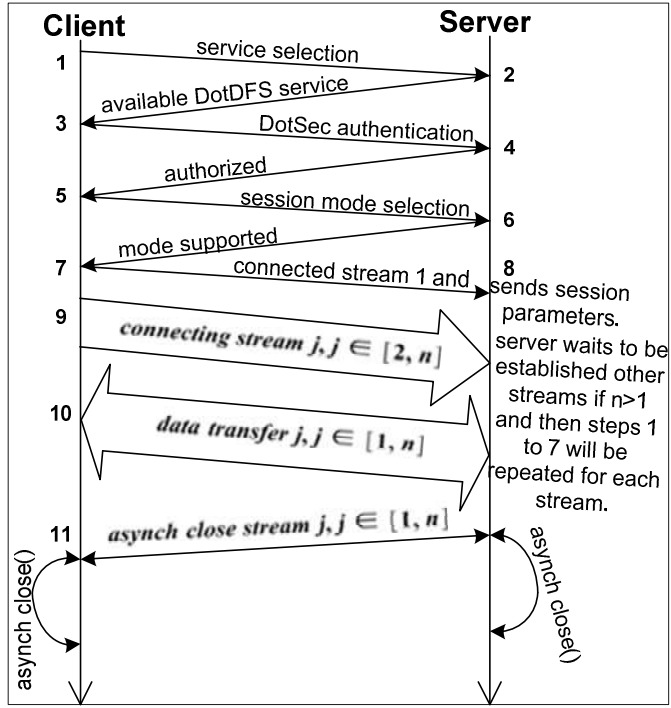

**Fig. 4.** Client-server DotDFS protocol communications in FTSM mode, *n* is the number of parallel streams.

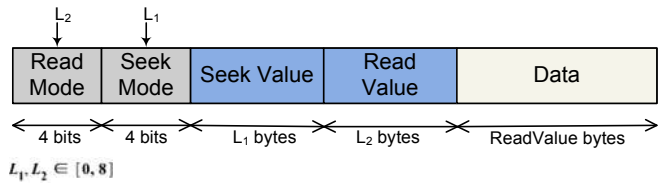

**Fig. 5.** Download/Upload binary header for FTSM mode.

Upon completion of session selection mode (stage 5 in Fig. 4), client sends FTSM mode request. If the supported mode notification header is submitted from the server, the first stream connection of client is now established and a new DotDFS session is created in server side. In this step, the client sends file transfer session parameters including the number of parallel streams, TCP Window buffer size, Transfer Type (Upload/Download), and a unique GUID.

In the next step, server is in waiting mode until all the specified parallel streams by *stream1* are established. The connection schemas of the remaining streams are similar to the first stream. When all of the parallel streams are established, due to the Transfer Type (Upload/Download), the bulk data transfer of stage 10 is started.

It is necessary to note that in both upload and download sessions, the stream connections are initiated from client side to server side. Furthermore, depending on the data flow the following operations are done in the client-server sides: reading from local storage and sending to the remote server (in upload scenario initiated by the client), and receiving from the remote server and writing to local storage (in download scenario initiated by the client).

In the DotDFS protocol, server does not connect to the client, and it seems this is a good feature in our proposed protocol. This results in omitting concepts like control channel used in GridFTP. This feature also solves some problems of GridFTP protocol addressed in [9,10,11]. These problems in more detail are discussed in section 2.4.4.

In Fig. 5, our suggested and implemented binary protocol headers during data transfer channels in stage 10 of Fig. 4 are shown. In Fig. 5, the SeekValue is the value of file offset for transferred file block and ReadValue is the transferred data length. These two blocks have variable length specified by $L_1$ and $L_2$ respectively. If both of $L_1$ and $L_2$ are zero, this is a notification of successful transfer of all file blocks. The connection side receives the zero value, closes the established streams and releases the used OS resources.

FTSM mode supports the feature of reusable transfer channels simply by adding the required headers shown in Fig. 5. These headers must be included during the DotDFS session initiation on the connection establishment.

## 2.3. PathM Mode

PathM mode is designed to support basic features like creation/deletion of remote files/directories and related features that are depicted in Fig. 1. Fig. 6 shows the client-server communication architecture of DotDFS protocol in PathM mode. Steps 1-7 in Fig. 4 are also used in Fig. 6. In the PathM mode, DotDFS server operates like a RPC server, but all the requested methods of client are previously defined as binary in the client-server negotiation protocol.

The PathM mode methods for client-server communication are based on the mechanisms shown in Fig. 1 and 6. PathM mode also supports a unique feature for upload/download of huge small size files in the client side. This is implemented via Upload/Download SmallFile methods.

The reusable channels support in PathM gains the RPC server-like functionality to our proposed protocol. In PathM mode each client may ask consecutive commands to execute from the server many times during one open DotDFS session. In section 4.2, a sample scenario of this functionality is presented for DotDFS Directory Tree Transfers.

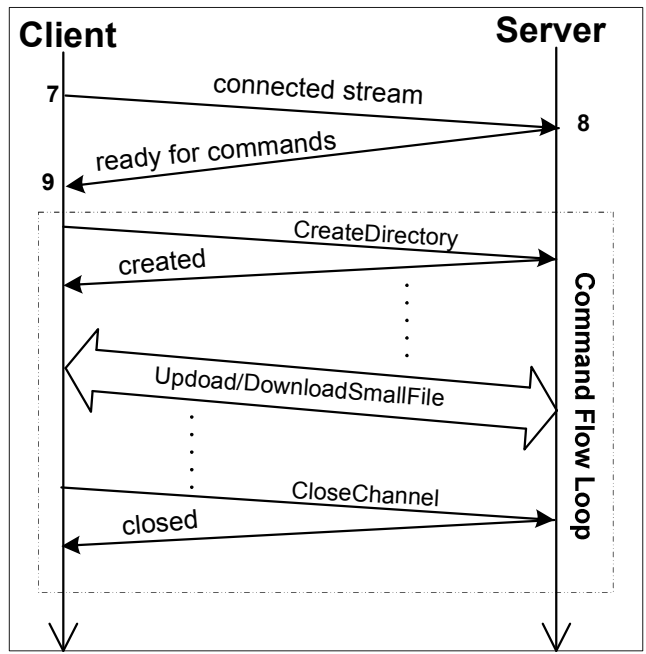

**Fig. 6.** Client-server DotDFS protocol communications in PathM mode.

*2.4. More DotDFS Protocol Features*

The DotDFS protocol includes many other features and specifications. In this section, we briefly state a few other features available at the current proposed protocol.

*2.4.1. Striped Data Transfers*

In FTSM mode, the TSI/X-Channels (shown in Fig. 7 and described in sections 2.2 and 4.1) may actually consist of several TCP streams from multiple hosts identified by GUID's that cause the realization of the striped data transfer support in DotDFS protocol. Data may be striped or interleaved across multiple servers, as in a parallel file system. As well as, stripped transfers is fully supported in DFSM mode [12,13].

*2.4.2. Partial File Transfers*

Partial file transfer helps Grid processes and applications tend to transfer only fragments of a file. This mechanism is supported by pre-defined DotDFS headers.

*2.4.3. Stateless Architecture*

DotDFS protocol introduces a natural stateless architecture in DFSM operating mode. This means that DotDFS servers do not keep track of DotDFS client requests according to opened files, file positions and etc. This leads DotDFS clients to be able to fail and resume without disturbing our system as whole and allows fast development of Grid-based parallel cluster file systems, in future.

*2.4.4. The Firewalls' Issue*

For intermediate solving diverse problems with the firewalls stated in [10,11], all DotGrid services are running on default listening port 2799 or client API calls are enforced to be established on this port. On each machine, one process called DotGrid Listener listens to this default port for Grid task requests. After receiving a service request to this process, based on the sent binary header request from client, the selected request is relayed to an appropriate service runtime manager like DotGridThread, DotDFS and DotGridRemoteProcess stacks. The remote service selection is done by DotGrid binary request header protocol. Then, as stated in the sections 2.1, 2.2 and 2.3, all upload and download requests in DotDFS protocol must be originated from client to the server.

**3. High-Performance Server Design Patterns for DotDFS Protocol**

Constructing highly concurrent systems is inherently sophisticated and difficult. This issue is more obvious in communication protocols where the protocol specification can greatly influence on the implementation and development methods of the protocol. While threads are the most commonly used tools to specify concurrency but large resource consumption and scalability limitations in many systems relied upon threading models have made researchers focus on event-driven techniques. Building a system merely based on threading models or event-driven methods is the main issue to make the system complex and difficult to evolve from different design aspects. A very good way for designing a concurrent system is to establish a level of balance between these two models. Event-driven techniques are advantageous to achieve high concurrency; but when systems architects take the problem of developing real time systems into consideration, threads bring out significant importance to exploit multicore-multiprocessor parallelism and deal with blocking I/O mechanisms appropriately. Since file transfer protocols like FTP and NFS are more disk I/O-bound, almost no research work has been conducted to suggest a concurrent file transfer protocol that simultaneously employs threaded and event-driven models in the protocol level. Also, the specification and implementation of legacy file transfer protocols nearly remain intact in the level of the processed model, multithreaded model or threaded pool models. Due to our knowledge DotDFS is the first concurrent file transfer protocol that, from this viewpoint, presents a new computing paradigm in the field of data transmission protocols. In the remainder of this section we examine the benefits and disadvantages of two these models and will finally introduce the hybrid server architecture for DotDFS protocol discussed in section 2.

### 3.1. Threaded-Based Concurrency Pattern (TBCP)

A thread of execution is the smallest processing unit that can be scheduled by an operating system. Operating systems usually allow processes to switch between different threads through thread scheduling primitives such as preemptive and cooperative scheduling; this operation is called the term context switch. A primary overhead exposed by threads is $O(n)$ in the number of threads. While the number of threads is increasing, the system experiences overheads, including scheduling and context switching, memory pressure due to thread footprints, cache and TLB misses, and contention for shared resources such as locks. In highly threaded systems, the instruction cache tends to take many misses as the thread's control passes through many unrelated code modules to process the task. When a context switch explicitly takes place by the OS preemption or cooperatively by the requested thread, other threads will invariably flush the waiting thread's state out of the on-chip cache.

### 3.2. Event-Driven Concurrency Pattern (EDCP)

As stated in the previous section, threads exhibit many scalability limitations in developing highly concurrent systems. Event-driven methods are used to overcome problems of threads. In EDCP model, a server consists of a certain number of threads that loop frequently and process events with different types from a queue. In this way, each request is described as a set of finite state machines (FSMs). Events are usually generated by operating system kernel and accessible through system calls in forms of network and disk I/O readiness, completion notification, and timers. As seen in TBCP model; the number of context switches decreases due to multiplexing requests in a definite amount of threads; avoiding synchronization mechanisms not only increase the performance but reducing the number of consumed threads strongly boosts instruction and data locality.

### 3.3. Hybrid DotDFS Concurrency Pattern (HCP)

As shown in two EDCP and TBCP models both have advantages and disadvantages. In this section, we propose a hybrid concurrency pattern for DotDFS protocol, which directly dictates the protocol implementation and specification. The main idea of this methodology is the use of an approach to balance between EDCP and TBCP models in favor of the HCP model. In this section, to show the efficiency of our proposed approach we only examine the structure of HCP model for DotDFS FTSM upload mode in the server as the wildly-used case in Grids.

Because DotDFS is a disk I/O bound protocol, to detach the protocol from non-standard asynchronous disk I/O interfaces for maintaining the protocol portability and universality features we have made use of threads to provide a nonblocking disk I/O architecture. Each thread manages one DotDFS session. To use the benefits of event-driven techniques, multiple TCP streams are processed through event dispatching methods for each thread of transfer session. Therefore, in the proposed architecture, threads are used to execute parallelism on multiprocessors in order to eliminate blocking disk I/O drawbacks, and event-driven techniques are used to exploit high-throughput data transfer through nonblocking network I/O interfaces. Considering the diversity and heterogeneity of the system functions that allow asynchronous network I/O to be used, our current focus on the event-driven component of HCP architecture has been given to the system function *select()*.

Fig. 7 illustrates a typical timing diagram of the event-driven DotDFS server architecture per each FTSM upload session in which *n* parallel connections are processed in a thread of execution. In this diagram, the infinite event dispatcher loop is the main part of the event-driven system that continuously checks the read-readiness of sockets in the black-border loop. After selecting *m* sockets by the event dispatcher loop, the finite I/O loop in an iteration with the size of *m* writes the packets received from the protocol subsystem through the system function *recv()* into the hard disk by calling the system function *write()*. The top side of the timing diagram shows mode switches (mode transitions), and the bottom side shows copy operations, between user and kernel space. As this diagram tells us for every move from step 0 to step 6, there to the number of $2 + 4 * m$ mode switches occur. Mode switches in operating systems are very similar to a simple function call in user space in which the return operation is done to the user space after jumping to the kernel.

On most architectures, the cost of these mode transitions consists of saving all (or some, or none) of the registers to the stack, pushing the function arguments to the stack (or putting them in registers), incrementing the stack pointer and jumping to the beginning of the new code. Hence the HCP model only slightly suffers from mode switch overheads, this negative cost is notably little and can be completely ignored against the overheads exposed by the TBCP model. It is necessary to note that the overheads of mode switches cannot be removed in any way because user space applications (such as GridFTP server) must call system functions in having access to the hardware functionalities provided by proper kernel interfaces.

While our concern in the initial phase of DotDFS implementation was to maintain its portability running on the most existing platforms through POSIX-compliant system function standards like *write()*, *select()* and *recv()*, this approach exposes extra overheads due to additional copies occurred between hardware, user space and kernel memory buffers (this issue is also exactly concerned with GridFTP server). These overheads can be eliminated by using zero-copy mechanisms, considering these techniques are beyond the scope of this paper. As the bottom side of Fig. 7 shows, $4*m$ copies happen in steps 3 to 6 for each iteration of the inner loop highlighted as the green border, including CPU and DMA copies.

We have applied a novel method in the design of HCP model. As earlier stated, we have used *select()* to maintain more DotDFS portability. But in practice we cannot implement a full DotDFS server core with one thread and one loop containing the *select()* function; because this method not only reduces the server parallelism, but also passing a large set of socket descriptors to *select()* significantly decreases the scalability of the developed server in high-traffic environments due to frequent copies during the transition from user space to the kernel and vice versa. Since using five parallel streams is normally common in Grid, for example, to upload large files; thus it is more reasonable that these network streams are processed in a separate thread through multiplexing network I/O techniques. Using different event-driven models with more capabilities, which ultimately make DotDFS protocol to have more sophisticated architectures, will be one of our straightforward future research works.

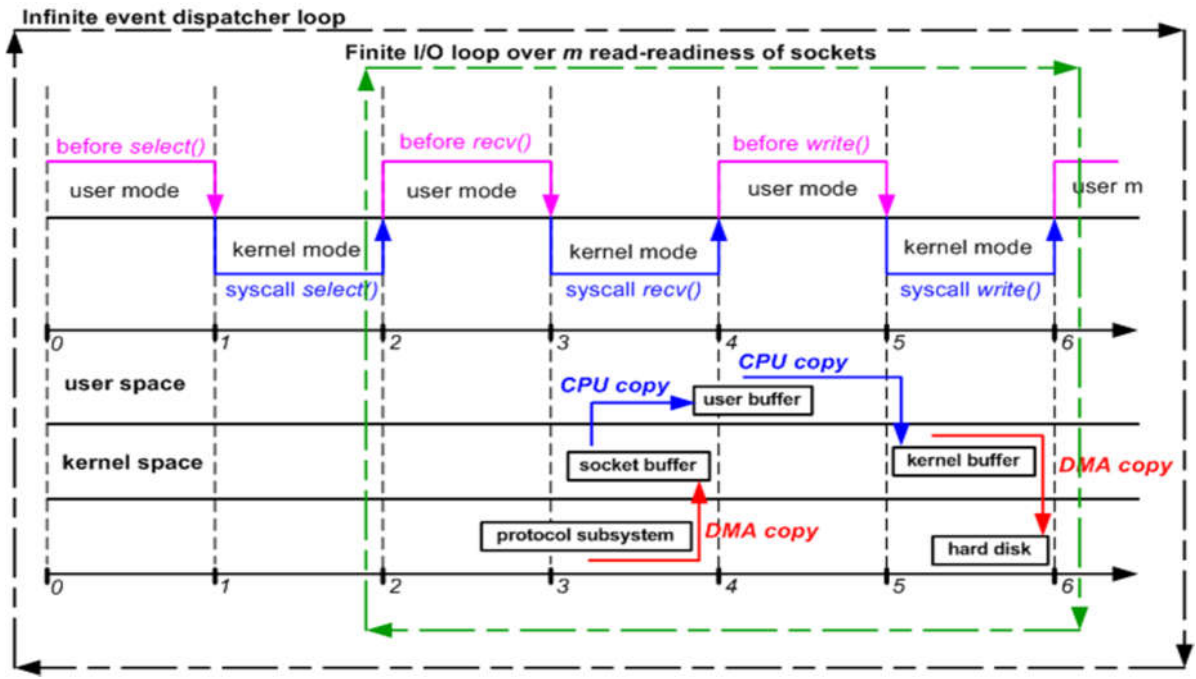

**Fig. 7.** A typical timing diagram showing the event-driven DotDFS server architecture per each FTSM upload mode's thread of execution.

## 4. Our Implementation of the Proposed DotDFS Protocol

This section presents an implementation survey of the main units of the DotDFS protocol proposed in section 2.

*4.1. Architecture of DotDFS Implementation in FTSM Upload Mode*

In this section, we describe how a DotDFS client in DFSM mode can transfer large file sizes to a DotDFS server by using parallel streams to achieve a high transfer rate. Fig. 8 shows the architecture of client-server DotDFS implementation in FTSM upload mode.

1. When a client wants to connect to a DotDFS server, the client calls the UploadClient API (the client has to declare some parameters such as the number of parallel streams, file size, TCP Windows size and the underlying DotSec TSI parameters when accessing to this client API). The UploadClient API creates a master thread in which management of *n* parallel connections is flowed. At the beginning of the session, this thread generates a GUID that will be used to distinguish all the counterpart parallel connections for a DotDFS transfer session in the server side. All streams ($stream\ j, j \in [1, n]$) are established by this thread to the remote DotDFS server.

We have implemented a Local File Read Queue (LFRQ) in DotDFS APIs to decrease side effects like extra local storage activity, opening large file handles for large parallel streams and random file access in using extensive *Seek()* file system calls. LFRQ sequentially reads file blocks from the file system and put them into a queue. A locking mechanism is used when the *stream j* issues a request to access a file block from the buffer. While a read request is being received to the queue, all other file blocks in LFRQ are locked to provide a coherency model until the request of file block read completed.

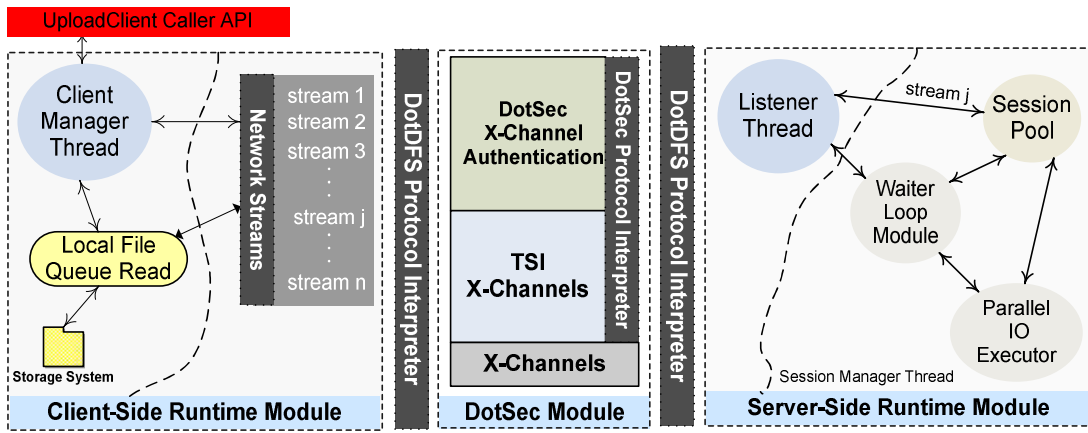

**Fig. 8.** Architecture of client-server DotDFS implementation in FTSM upload mode.

The throughput of file block read from LFRQ is directly depended on available external bandwidth and network latency in which the DotDFS session is being executed via *n* parallel streams. With regard to this approach, there are always enough file blocks provided by LFRQ for the requesting streams. After the complete establishment of a DotDFS session, the master thread managing the DotDFS session investigates writability of sockets, which constitute the endpoint streams, by successive calling the system socket *select()* function. In this time, the master thread performs requesting the needed file blocks to the number of selected sockets by the system *select()* function from LFRQ and sends them over the writable streams. All opened file and socket handles will be released and returned to the host operating system at the end of each DotDFS session (it means that all file blocks had been sent to DotDFS server) by the manager thread.

2. The following operations are performed at DotDFS server in each DotDFS session. A thread listening on default port 2799 receives the requests from the network adaptor. When a request is received, this thread examines the sent parameters by the *stream j*, and extracts the GUID parameter which the stream contains. The listener thread investigates whether the GUID exists in the session pool's hash table or not. If there is not any GUID related to the received GUID, then it represents the starting of a new DotDFS session. In this state, if *n* (the number of parallel streams) is greater than zero, then the listener thread creates a secondary thread, named as Manager Thread (MT), and invokes Waiter Loop Module (WLM) method, which contains in the class's member method that is executed by MT (also parameters like *n*, remote file name, local file name and TCP Window size are passed to this module).

Also, MT puts the received GUID into the session pool. In next requests that are received by the listener thread, if these requests' GUID is existed in the session pool then *CurrentReceivedStreams* variable related to MT is increased with value of 1 by the listener thread. Since the beginning of the execution of WLM by the listener thread, WLM is continuously checking the value of *CurrentReceivedStreams* variable and comparing it with the passed value of *n*. When *CurrentReceivedStreams* equals to *n*, WLM terminates its loop iteration and runs Parallel IO Executor (PIOE) module, while deleting references to the counterpart connections related to the current session specified by its GUID.

PIOE is the main core of the DotDFS server. POIE in an infinite loop, similar to the mentioned client side model, investigates the readability of sockets, which constitute the endpoint streams, by successive calling system socket *select()* function. POIE selects the readable sockets, receives file blocks from these sockets and writes them into the storage system. Because networks like WAN usually have high latency behavior, we cannot consider any certain coherency from the aspect of file block offset for receiving the sent file packets from client to server side. When file blocks are conveyed based on DotDFS protocol, a set of information is sent to the destination server encapsulated in the headers of DotDFS protocol by DotDFS client, such as the block offset and block count. Hence in regard to the random receiving of these file blocks, POIE must call *Seek()* file system function in a repetitive sequence. To decrease this overhead, we have implemented a queue module similar to the LFRQ in the DotDFS client side. It attempts to queue file blocks, that are located in an offset interval, and write the whole queue capacity into the storage system. The overall effect of this technique is that the deteriorated coherency can somewhat be increased, and consequently, decreasing the *Seek()* system calls. In this way, we make use of buffering the new received file blocks in a typical 1MB buffer for coherency preservation and decreasing random file access through *Seek()* file system calls with writing the total buffer to the storage system. At the end of each DotDFS server side session, all consumed operating system resources will be released like opened native file handles and threads by session thread manager.

In the above mentioned paragraphs, only a total of two threads are created in client and server sides that are responsible to manage a shared DotDFS session. This fact implicitly explains that the optimized kernel operating system methods in the proposal and implementation of the DotDFS protocol have been considered.

*4.2. DotDFS Directory Tree Transfers*

There are many situations where moving a directory tree is desired. For example, a grid user may wants to migrate from his/her whole home directory to another machine for load balancing or capacity-limit reasons. The DotDFS client and server architectures are extended to gain the directory tree transfer using the FTSM and PathM modes. The new implemented APIs are then included in DotGrid Platform SDK. The overall view of operations is illustrated in Fig. 9.

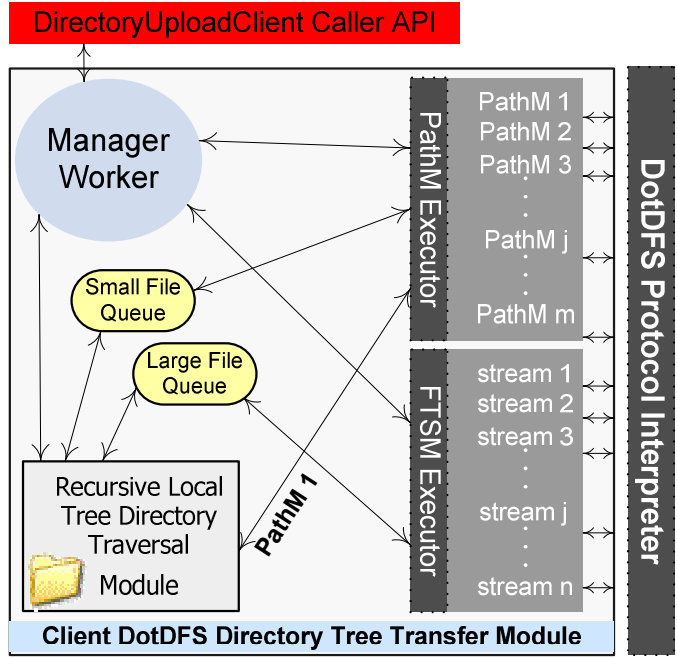

**Fig. 9.** Client DotDFS architecture for directory tree transfers.

In Fig. 9, the upload scenario of a local directory from client to a DotDFS server is depicted. When the client calls DirectoryUploadClient API, the following tasks are done and managed via the Manager Worker (MW):

1. First, MW establishes the *PathM 1* route to the server. *PathM 1* is the operating agent of recursive creation of the remote directory tree. Moreover, MW runs the Recursive Local Tree Directory Traversal Module (RLTDTM). Not only RLTDTM is the responsible agent for remote directory tree creation, but also it queues the found files into Small File Queue (SFQ) and Large File Queue (LFQ) modules based on small or large file sizes in each state of its directory traversal.

2. After completion of the local directory traversal and creation of the remote directory tree, MW creates *m-1* connections (*m* is the number of parallel connections in PathM mode) to DotDFS server. Files queued in SFQ are transferred between two endpoints through these established channels. Also, MW simultaneously calls DotDFS UploadClient APIs stated in section 3.1 to transfer large LFQ files by using *n* parallel streams.

As seen in this scenario, *m+n* transfer channels are established between two endpoints in parallel. If the LFQ and SFQ are not empty and LFQ contains minimum *m* queued files, then *m+1* files are being transferred between a DotDFS client and server in parallel. In this approach, PathM and FTSM executor threads are in competition and do transmit SFQ and LFQ fetched files. There is no default compression algorithm in our implemented APIs. But these APIs have interfaces that a Grid/Cloud user may extend his/her own APIs with compression algorithms like GZIP, LZ77, BZip2, etc enabled.

The above example illustrates the usefulness of the layered model of implementation in DotDFS protocol, and features like multi-level parallelism and reusable channels mentioned in sections 2.2 and 2.3, for gaining high-throughput and high-performance directory tree transfers.

**5. Comparison of DotDFS Protocol with GridFTP Protocol**

In this section, we compare DotDFS protocol with FTP and GridFTP protocols. Some GridFTP weaknesses and structural differences between DotDFS and GridFTP are discussed, so that the reader can understand more reasons (in addition to those considered in sections 2, 3 and 4) why a new concurrent file transfer protocol is proposed. The aim of this paper is not to decline the GridFTP protocol; rather we will show when a new protocol is designed from the ground-up for the next generation of distributed computing paradigms how it can lead to different and better results from a completely new and deep perspective. However, Globus in [] expresses the following phrase for selecting and extending the FTP protocol: "We chose the FTP protocol because it is the most commonly used protocol for bulk data transfers on the Internet and of the existing candidates from which to start (http, DPSS, HPSS, SRB, etc.) ftp comes closest to meet the needs of Grid applications." []. Table 1 shows a comprehensive comparison of different file transfer mechanisms including DotDFS, FTP, and GridFTP.

TABLE 1
COMPARISON OF DIFFERENT WIDELY-USED FILE TRANSFER PROTOCOLS

| Feature | DotDFS | FTP | GridFTP |
|---|---|---|---|
| Creation year | 2010 | 1971 | 2003 |
| Protocol standards | DotDFS v.1 (2010), xDFS v.2 (2010) | 20 RFCs | GFD-R-P.020 (2003), GFD.47 (2005) |
| Low-Level Transmission Protocol | Multi-protocol support via DotGridSocket Interface | TCP/IP only | Multi-protocol support via Globus XIO |
| Platforms | Cross-platform | Cross-platform | UNIX/Linux |
| POSIX-compliant I/O standard support | Fully | No | Partially |
| Local-area file access support (e.g. NFS) | Fully | No | No |
| WAN improvements (TCP window size/parallelism) | Strong/Strong | No/weak | Strong/Moderate |
| Native event-driven protocol-level architecture | Fully | No | No |
| Statefull/stateless architecture | Fully/fully | Fully/No | Fully/No |
| NAT compliance | Strong | Weak | Weak |
| Firewall compliance | Strong | Weak | Weak |
| Internationalization | Strong | Moderate | No |
| Extensibility | Strong | Weak | Moderate |
| Protocol message interchange exchange format | Binary | ASCII | ASCII |
| Security extensions | DotSec | 7 RFCs | Globus GSI, GSS-API, and FTP RFCs |
| Scalability/Large concurrent requests support | Strong | Weak | Weak |
| File/pipe-based inter-process communication support | Strong | No | No |
| Large size file support | Strong | Weak | Strong |
| Protocol-level zero-copy extensions | Yes | No | No |
| Distributed file systems semantics | Yes | No | No |
| Recursive Directory tree transfer support | Strong | Weak | Moderate |
| Operating System Recourse Consumption | Low | High | High |

## 5.1. DotDFS Protocol vs. GridFTP Protocol in Upload Mode

In this section, we concentrate on comparing DotDFS and GridFTP protocols. Since the complete comparison of these two protocols is out of the number of pages of this paper, we consider just a specific taxonomy to explain. In this scenario, we assume that a client wants to upload a large file to a server via *n* parallel connections. DotDFS and GridFTP protocols use FTSM and passive X PUT modes, respectively. The server-side CFSMs of these two modes are illustrated in Fig. 10 and 11. These CFSMs are concurrently representing the protocol implementations and largely the protocol specifications as a collection of FSMs. The red boundaries mean those parts are executing in parallel.

A suitable model for describing communication protocols and concurrent systems are communicating finite state machines (CFSMs). In the CFSM model, a protocol is defined as a collection of processes (i.e., the protocol entities) which exchange messages over error-free simplex channels. Obviously, the CFSM model virtually has become the de-facto standard to specify, verify and test communication protocols in the telecommunication industry.

DotDFS CFSM, to a large extent, is representing the descriptions discussed in sections xx and yy. After authenticating the client through DotSec GSI, choosing the FTSM mode by the client, and receiving the FTSM parameters; the server checks whether the session has already been created using its GUID by the client or not. If the session has already been existed and the number of sockets in the hash table is not corresponding (equal) to the value of *n* received from the client, the server adds the new client stream to the hash table in state 8. In step 7, the server concurrently checks, so that if the number of client streams is equal to the value of *n* then moves the CFSM flow to state 9. If an error occurs during states 1 to 8, the next state will be 15. States 9 through 14 are the main DotDFS server core where file blocks are received from the client only in a loop in one thread of execution, through network I/O event dispatching mechanisms, and the server writes them to the hard disk in state 12.

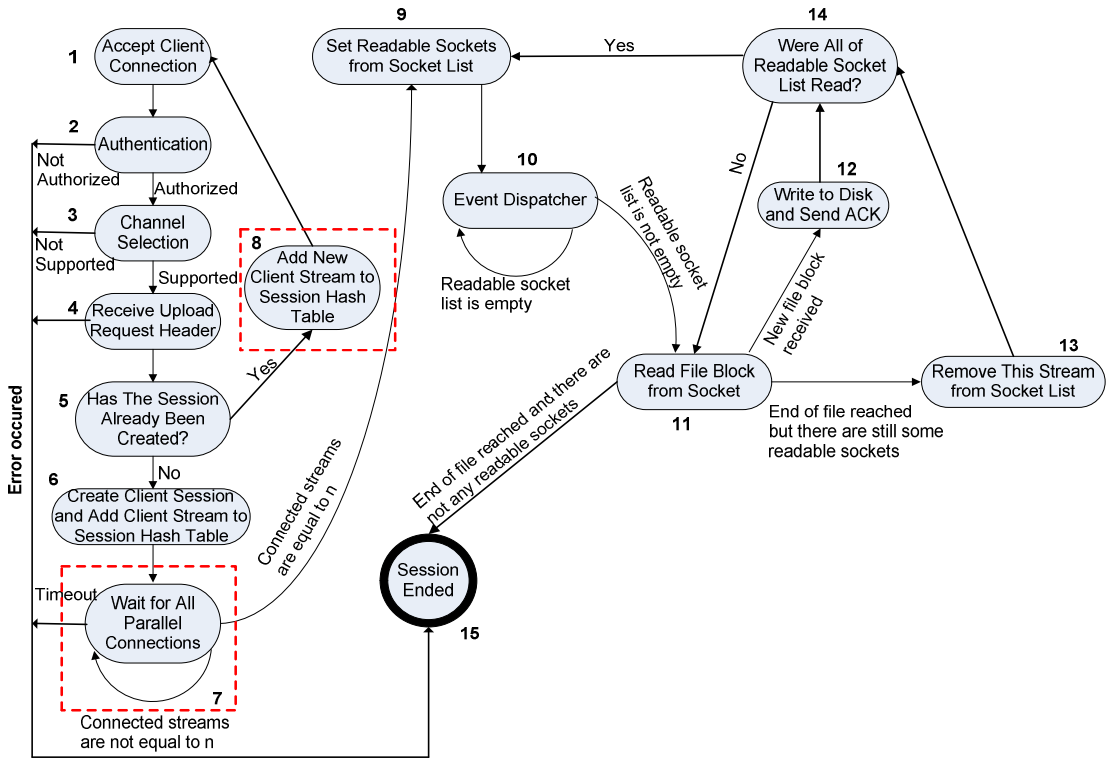

**Fig. 10.** DotDFS server communicating finite state machine in FTSM upload mode

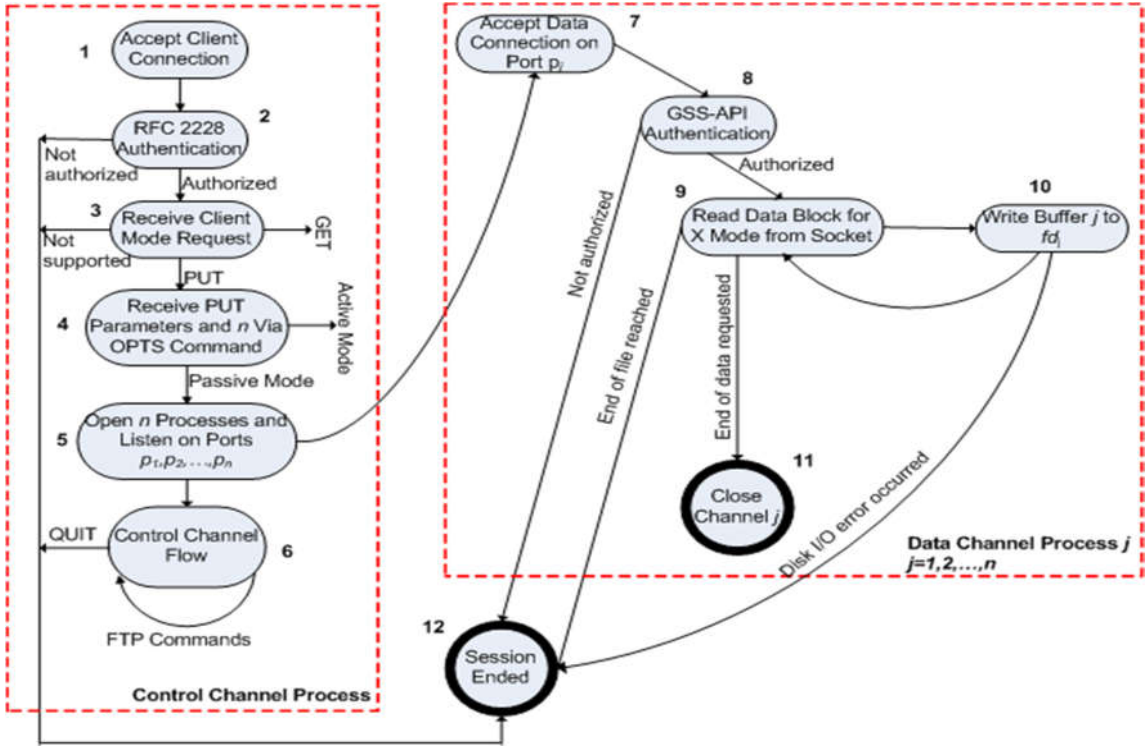

**Fig. 11.** GridFTP server communicating finite state machine in passive X PUT mode.

As the CFSM of the GridFTP server core shows (Fig. 11, inferred from GridFTP drafts and Globus published papers), it consists of $n + 1$ processes that contain the control channel process and data channel processes. GridFTP protocol in this scenario cannot use event dispatching mechanisms to open only one process; because, firstly the control channel and data channels are two separate modules that must be executed in parallel, and secondly the different types of the Descriptor field in the X mode header (such as EOR, EODC, EOD, etc.) do not permit the server to uniformly manage *n* sockets in a loop with event dispatching mechanisms used.

Additionally, in the source code level, Globus Toolkit makes extensive use of the UNIX *fork()/exec()* system functions to implement GridFTP client and server programs, and IPC methods to implement synchronization and communication between the control channel process and *n* data channel processes. Section zzz will argue the overheads of GridFTP implementation that are due to using these two functions. In this CFSM, the GridFTP server remains waiting for the client to choose PUT or GET mode after accepting the client connection and authenticating the client according to the RFC 2228. In addition to the parameters of the PUT command, the server in state 4 receives the number of parallel connections via the OPTS command. In state 5, the server creates *n* processes and remains waiting on ports $p_1, p_2, ..., p_n$ to accept ongoing client connections which will constitute the data channels. In state 6, GridFTP server receives the client requests as FTP commands in an infinite loop over the control channel. This allows the client to be aware of the transfer status (such as performance monitoring commands during X mode sessions that are sent and received by the client over the control channel) and to control the server-side GridFTP session over data channels. The server in state 7 accepts data channel connections and authenticates them through GSS-API mechanisms. At this stage, a separate file handle is opened and a memory buffer with the length of block size is allocated to perform the receive operation from socket and the write operation to the hard disk, for each data channel.

The above description reveals other facts during each server-side GridFTP session with the passive X PUT mode in which *n* file handles, $2n + 1$ socket handles (*n* socket handles to listen on *n* ports, *n* socket handle for accepted connections belonging to data channels, and a socket handle for control channel) and a discrete chain of buffers with size of $n * BlockSize$ are assigned. As it can be seen, the GridFTP implementation suffers from these overheads due to the inherent protocol nature. In contrast to GridFTP, DotDFS only requires assigning one memory buffer with the size of block size, opening a file handle and *n* accepted socket handles.

## 6. Lightweight DotSec's Grid Security Infrastructure Model

Virtual organizations share different and wide range of resources distributed in Grids. This urges the need for more security issues in developing Grid Computing [21]. In DotSec protocol, we propose a model of new lightweight grid security infrastructure. It has the security structure like Globus GSI [22] and SSL (as well as TLS [23]), and also provides some other security requirements for DotGrid services.

Recently, the Globus GSI has being played a key role in the security of Grid Computing. Mutual authentication and delegation, and single sign-on are major services of Globus GSI. Based on Public Key Infrastructure (PKI), Globus GSI clients must manage and process long-term credentials. PKI-based GSI is somehow a heavyweight approach. This is mainly due to public key certificates and proxy certificates [24].

This section is about the overall view of our new lightweight GSI solution of DotSec. DotSec is a lightweight GSI but not Globus GSI. It is designed as a unified security model in DotGrid platform. It uses extensively from all protocols and methods described here. The layered DotSec model is shown in Fig. 12.

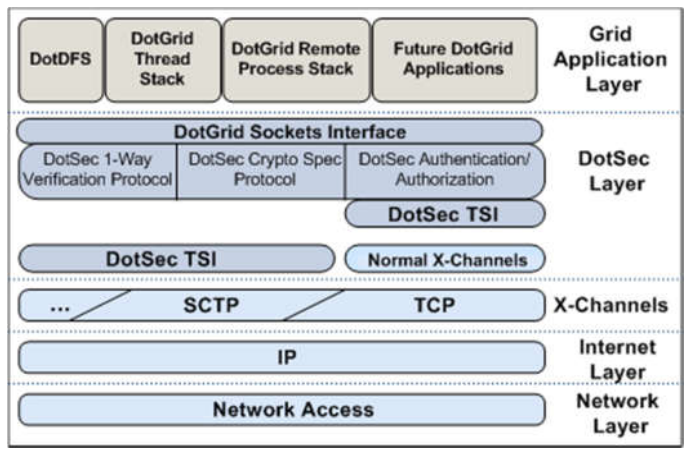

**Fig. 12.** Layered DotSec grid security infrastructure model.

As seen in Fig. 12, the DotSec layer not only supports Grid-based authentication/authorization mechanisms, but also includes TSI core, a critical service of data transfer security like SSL and TLS. DotSec also reduces the heavyweight interface of the most common protocols used in Globus GSI and SSL. It suggests services like change between TSI X-Channel state to Normal X-Channel and vice versa at each point of the transfer session among endpoints.

The private and public RSA keys import/export as Authenticode X.509 v.3 certification was added to gain more flexibility in DotSec protocol. The main DotSec core has a 4-layers structure and all the communications between services, clients, and DotGrid APIs are conducted through DotGridSocket API into DotSec sub-layers.

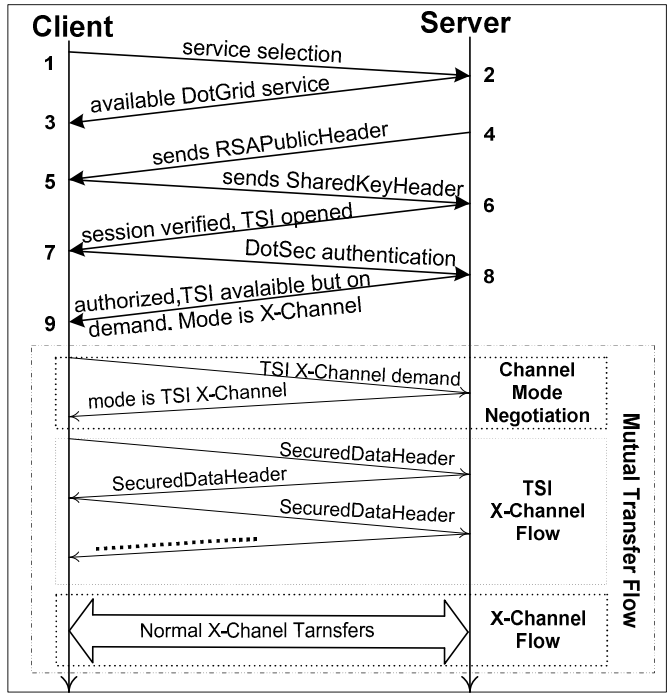

**Fig. 13.** Client-server DotSec layer protocol communications.

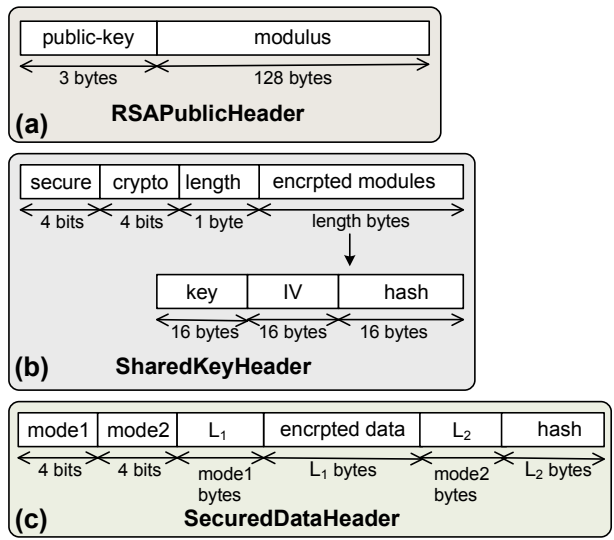

**Fig. 14.** DotSec protocol headers.

*6.1. DotSec One-Way Verification Protocol*

In the proposed DotSec protocol, the sharing of a private shared key is done for twofold purposes: the first is for using it in encryption tasks of all data transfers between two endpoints over TSI layer, the second is to verify and certify the clients connected to the server.

In Fig. 13, a DotSec session between a client and server based on our proposed DotSec protocol is illustrated. As shown in Fig. 13, the client waits to receive RSAPublicHeader from the server just after service selection of stage 3. The RSAPublicHeader structure is depicted in Fig. 14.a. It contains both of server public key and RSA encryption modulus.

Just after RSAPublicHeader receives, client manages the following tasks: first a private key and an initial vector (IV) are generated relied upon a symmetric cryptographic (SC) algorithm, second the hash of *key+IV* is calculated based on a hashing algorithm. And finally by using the public-key and modulus of RSAPublicHeader, all the data are encrypted and ready to submit to the server. The final data structure is shown in Fig. 14.b (the value of *key+IV* is used for data encryption between two endpoints by TSI).

In the next stage, server manages the following tasks: first it decrypts the received data using its own RSA private key, second calculates its hash, and finally compares the calculated hash with the hash in arrived SharedKeyHeader. The client is certified and verified only if these two values meet.

The certification/verification stage initiates the TSI between client and server. This in turn allows to use client submitted shared key for encryption if there is a need for secure data transfer in DotSec protocol or if there is a request for whole encryption in the DotSec sessions.

*6.2. DotSec Authentication and Authorization*

After client verification, client submits the username, password and all the other requested service parameters (e.g., remote path in DotDFS PathM mode) using TSI as encrypted with shared key. In the server side, based on the submitted data and resource access control, the user is authorized if it has been validated and the communication session has already legally been started.

DotGrid permissions service is described in [12,13] and presents the way in which a client is allowed or not to use the provided Grid resources during DotGrid and DotSec sessions.

*6.3. DotSec Cryptography Specification Protocol (DotSec CSP)*

DotSec CSP is used for selection and negotiation of Symmetric Cryptographic (SC) and hashing algorithms used for shared key generation and encrypting the data sent over X-Channels.

As shown in Fig. 14.b, the *crypto* block in SharedKeyHeader is indicating the type of used SC algorithm in the client side toward the server side and the server must validate and acknowledge the demanded SC type based on DotGrid service provider's administrative rules.

In the current version of DotSec, the following algorithms are supported by default: SC algorithms including Rijndael proposed by NIST, Triple DES (using one key and IV for encryption), and hashing algorithms of SHA-1 and MD5.

*6.4. DotSec Transmission Security Interface (DotSec TSI)*

TSI is responsible for data encryption and transfer using shared key between two endpoints of the connections. TSI transfers the encrypted data encapsulated as SecuredDataHeader blocks (illustrated in Fig. 14.c) which contain important information about the encrypted data. Data integrity is assured through TSI in the following manner. As shown in Fig. 14.c, the hash block is the hash value of unencrypted pure data. It is used for verification of transferred encrypted data block. Further, this is used to assure whether the data in the DotSec session is intentionally tampered or not.

If the values of *mode1* and *mode2* in SecuredDataHeader are zero filled and submitted, then the receiver changes the working mode from DotSec TSI X-Channels to normal X-Channels shown in Fig. 13. This is a new feature in DotSec TSI protocol. It means that in this new mode, the transferred data is not encrypted and directly placed on X-Channels layer from DotGridSocket interface. We called this new mode in TSI as semi-secure.

As illustrated in Fig. 13, from step 9, user or DotGrid APIs' developer in use, extending or designing of new services for integration with DotGrid must set to enable/disable the mode of TSI in each step of data transfer explicitly with the structure of mutual transfer flow shown in Fig. 13.

## 7. Experimental Studies

Our current implementation of the DotDFS protocol includes all the features and specifications mentioned in this paper. For test and evaluation of our proposed protocol, some sort of experiments are prepared and tested. The results of our experiments show the effectiveness of the proposed protocol. Our experiments are done in three categories: LAN, WAN and directory tree transfer experiments.

Our LAN platform was a network with 1Gb/s and 0.05 ms RTT. The host machines of both client and server had Intel Dual Core 3 GHZ CPU, 1GB RAM, 360GB RAID hard disk. The target operating systems were CERN Scientific Linux 4 for x86_64 with 1G byte swap space. In LAN experiments, the TCP buffer size was set to 4MB. The directory tree transfer experiments are described in section 5.3. The WAN experiments are done between two machines with Windows Server 2003 placed apart on Miami, FL and Los Angeles, CA. The target platform was a network with 20Mbit/s and 76 ms RTT measured by Iperf tool.

Current version of DotDFS protocol is implemented on the top of DotGrid platform based upon Microsoft .NET Framework 1.1. We fully ported the developed toolset on Linux via MONO .NET Framework. This gains our tools to be portable on UNIX and Solaris clones too. In our experiments, Windows machines were with Microsoft .NET Framework 3.5 installed and Linux Machines were with MONO .NET Framework 1.9.1 installed. In our LAN experiments, the developed DotDFS protocol is compared with the Globus GridFTP protocol. We used Globus GridFTP of GT4.2.1 and GT4.2.1's globus-url-copy (GUC) GridFTP client utility [4].

The current version of the implemented DotDFS client enriches with a set of APIs for Grid developers. A new command-line DotDFS client utility with the most features of Globus GUC is also developed. In all experiments, the TCP channels' authentications in DotDFS are based on DotSec and while in GridFTP are based on GSI (GSS-API Grid Security Infrastructure) with Kerberos. All data points are the means of 10 runs.

*7.1. Single Stream Performance in LAN*

In the first experiment, the evaluation of DotDFS versus GridFTP in a single stream over LAN is done. Fig. 15 shows the performance and throughput for transferring files with size of ranging from 200MB to 4000MB. While our current implementation is suffering from the overheads of MONO .NET implementation and .NET Platform Invoke calls (more meaningfully due to the virtual machine nature of the .NET Framework), the proposed DotDFS protocol has reasonable performance over GridFTP. An interesting observation is the reduction of throughput for files larger than 1GB. The measured throughput in this state is rather much less than the speeds of the local read and write over the storage system. This phenomenon is discussed with more detail in section 5.2.

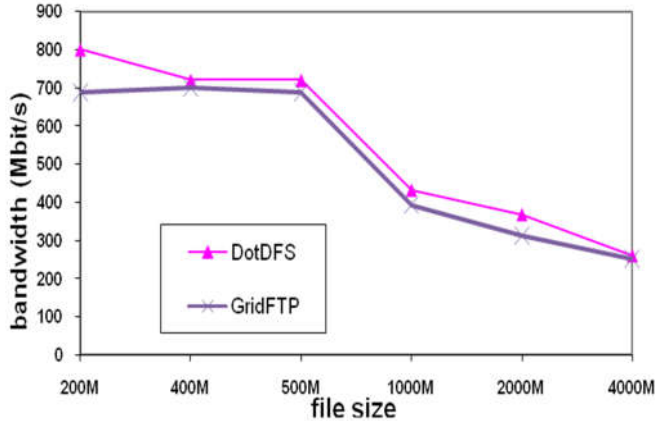

**Fig. 15.** Single-stream throughput on LAN.

## 7.2. Harnessing Parallelism in LAN

In the next step, we experiment and evaluate the effect of parallel TCP connections to increase the throughput. Fig. 16 shows the results for large set of streams. Fig. 13 shows three distinct evaluation modes: memory-to-memory (/dev/zero to /dev/null), disk-to-disk and Iperf test. In disk-to-disk tests a file with 1GB size is transferred between a client and a server. In the memory-to-memory tests, DotDFS and GridFTP reached the 94% and 91% of the bottleneck bandwidth, respectively.

As seen in Fig. 16, with increasing the number of parallel streams from 50 to 500, the memory-to-memory throughput is rapidly reduced in Globus, while is nearly constant in DotDFS. This phenomenon is more discussed in this section.

In our belief, the DotDFS results in section 5.1, disk-to-disk and memory-to-memory for parallel streams are not manifesting the real performance of our proposed DotDFS protocol. We think this is due to the overheads raised from using .NET Platform Invoke technology of .NET Framework in our current DotDFS implementation.

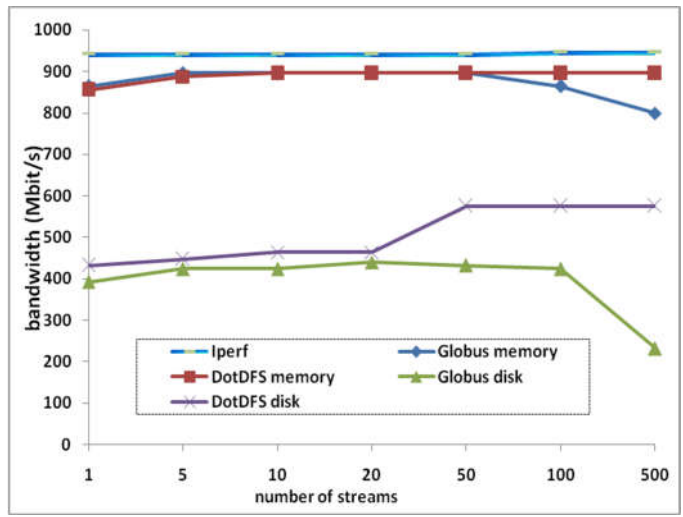

**Fig. 16.** Parallel throughput on LAN.

.NET Platform Invoke is a technology for calling native functions relied on metadata to locate exported functions and marshal their arguments at run time [25]. The .NET Platform Invoke technology is built right into the CLR runtime to enable managed programs (means the CLI runtime) to invoke ordinary dynamically linked unmanaged code. When working with unmanaged code, whether it is a native system function or native libraries in C++, there is a type system gap that must be bridged. For instance, a string to .NET Framework is not the same thing as a string in C++. Marshalling performs the transformations to the bits such that data instances can be used on both sides of the runtimes (managed runtime against unmanaged runtime). This operation may be a simple bit-for-bit copy from the managed to unmanaged runtime and vice versa, but just as well might involve a complete reorganization of the contents of a data structure as the copy occurs. This translation adds extra overheads. Hence, the marshaling mechanism can be very expensive; it can add tens of native instructions per argument for even simple native function calls. Calling native socket (through current DotGridSocket interface), file system and threading APIs by .NET Platform Invoke expose some overheads and shortcomings on our current DotDFS client and server implementations.

To unveil this deduction, we also implemented two other applications for data transfer between client and server in a memory-to-memory scenario with native standard C code and the .NET Framework based on C# language. Our new experiments show that the socket system calls in .NET add at least an overhead of 5% to our system.

This means that if DotDFS is implemented in native code by C or C++, it may reach about 99% of the performance and throughput of Iperf standard in the memory-to-memory scenario as well as the maximum performance in speed for read from the file in the sender and write to the file in the receiver in the disk-to-disk scenario (again, this means that if the current DotDFS implementation is ported to naive code rather than Microsoft .NET, one can access performance of benchmarks like Bonnie file system benchmark [26] with the notion of a high-throughput file transfer system).

As seen in Fig. 16, the disk-to-disk performance of DotDFS is better than Globus. By increasing the number of parallel streams (i.e. 500), the Globus throughput is reduced in disk-to-disk and memory-to-memory tests while DotDFS throughput remains remarkably high.

The throughputs shown in Fig. 13 promote that our DotDFS protocol is more suitable than Globus GridFTP for high-throughput data transfer. Globus GridFTP server widely uses from Linux forks instead of kernel threads like POSIX threads for multitasking server architecture support. This is specially used in parallel TCP streams implementation proposed by GridFTP protocol.

Threads are lightweight processes while forks are heavyweight processes. This explains that using UNIX forks in server-side results in higher overload and cost over using threads. This, in turn, causes more unstable server applications in dynamic, high traffic and actual server environments. The second overhead factor of GridFTP is the context switching. A context switch is the switching of the CPU from one process or thread to another in operating system or hardware level. Context switching is generally computationally intensive.

GridFTP protocol supports parallel TCP connections by the command PORT and other GridFTP protocol extensions. Fig. 17 shows an example of GridFTP passive file upload scenario in X mode. In this example, the GridFTP server replies on PORT command and go to listening state to accept new data channels for establishing parallel TCP streams.

```
Client            Server
PUT path=/tmp/file.dat;pasv;mode=x;
          1xx wait
          1xx wait
          1xx PORT=134,23,145,2,48,114
          1xx Data connection established
          2xx Transfer complete
```

**Fig. 17.** GridFTP passive file upload scenario in X mode.

According to this scenario, in GridFTP protocol, it is mandatory to use multi-threading or forks for parallel data channels establishment on the GridFTP server. Furthermore, the number of client-side requested threads and or forks are equal to the number of parallel data channels plus the one thread or fork for the data control channel.

This action results in more context switching (i.e. in large parallel streams or large clients' connections) in Globus GridFTP server. On the other hand, context switches for threads are faster than Linux forks, while Globus GridFTP extensively makes use of forks.

As previously stated in sections 2.2 and 3.1, in DotDFS protocol, for example, in an upload from client to the server, there is only one openen thread in the server-side, and data and session managements are all proceeded by this thread using consecutive socket *select()* calls in parallel TCP channels. This approach is somehow like the Single-Process Event-Driven (SPED) model [27].

This may describe clearly why the DotDFS throughput in disk-to-disk and memory-to-memory remains fixed with parallel streams increased while reduced in Globus GridFTP.

This observation leads to an interesting real world solution. In practical situations, for example with 500 concurrently connected clients and 5 parallel streams (which is widely used in WAN), then DotDFS server needs 500 threads and GridFTP server needs 2500 forks. This practically leads to more stable servers developed based on DotDFS protocol implementation.

To show the effect of more GridFTP protocol overheads, the percent of used physical memory in both client and servers of DotDFS and GridFTP servers are measured with the increase in parallel streams. The results are shown in Fig. 18.

As seen in Fig. 18, the rate of rise in Globus is much more than DotDFS by increasing the number of parallel streams. The main responsible factor for this increase in Globus is the more opened forks and used memory buffers for GridFTP transfer sessions when parallel streams are increased.

In our experiments, we also observed an interesting phenomenon. In all of our experiments for LAN, we observed that with the increase of file sizes more than 1GB, the performance and overall throughput are reduced with a factor and reached to a fixed value which is below the real measured speed of read and write on the hard disks used for client and server machines. This shows saturation on throughput. We call this speed as saturation speed. Saturation speed is achieved earlier with file sizes like 4GB and with increased parallel streams. We think that the main reasons for this observation are: first is data copies via reads and writes between user space and kernel mode for file descriptors and sockets, and second is the scheduling and context switching overheads due to the event handling and notification via socket *select()* calls in DotDFS or Linux forks handling in Globus GridFTP.

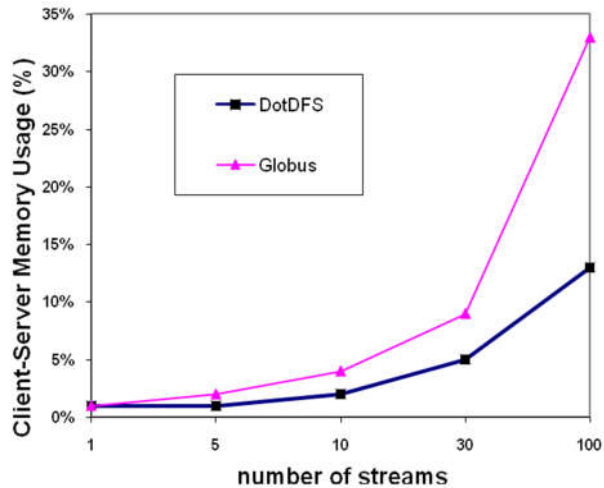

**Fig. 18.** The percentage of total client-server memory usage.

## 7.3. Transfer Large Collection of Small Files

In another experiment, we evaluate the performance of directory tree movement by DotDFS and Globus GridFTP. The directory to transfer contained 1000 files in 100 sub-directories. File sizes range from 1KB to around 512KB. The total data volume size in the directory tree is about 80MB. The achieved results are depicted in Fig. 19. For Globus case in GUC, we used the *–pp* option that provides GridFTP Pipelining which improves the transfer performance of lots of small file [28].

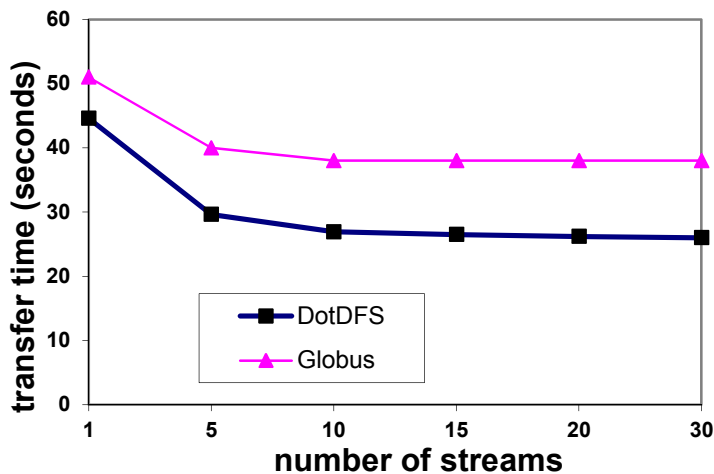

**Fig. 19.** Comparison of directory tree transfer performance.

As stated in section 3.2, DotDFS for directory transfers uses the full parallelism features in our proposed protocol and the efficient paralleled client side implementation. These are the main reasons that with increasing the number of streams, the whole performance in the directory tree movement is improved significantly in the DotDFS case rather than the GridFTP case.

## 7.4. WAN Experiments

In this section, a new idea for description and modeling of data transfer protocols in TCP/IP stack based on WAN's is presented. Due to our limited access to various topologies of MAN and WAN networks, the detailed development of this model is in our next tasks. There have been a large number of studies in the area of buffer optimization and parallel stream optimization. However, a comprehensive combination of tuned buffer size and parallel streams can even give more effective results than the single applications of these two methods. Unfortunately, there are not any practical approach to counterpoising the buffer size and the number of parallel streams to achieve the optimal network throughput.

In WAN, there are various facts including distance, TCP window size, packet loss due to the TCP congestion, and random network packet losses that mainly affect overall throughput of a single TCP flow. Furthermore, the lack of tuned TCP options also drastically reduces the throughput and wastes the useful bandwidth.

Many researchers have investigated to increase the effective WAN throughput (WAN goodput) especially in networks with high bandwidth-delay products and random losses.

In [29], the authors present a simple stochastic model of steady-state TCP behavior in its congestion-avoidance mode. This model states that the well-known result for a single TCP stream throughput is formulated in equation 1 of Table 1. In this equation, *MSS* is the maximum segment size, *RTT* is the round trip time, *C* is a constant term that reflects details of the TCP acknowledgment algorithm (*C*=1.22 if every segment is acknowledged and *C*=0.87 if at least every other segment is acknowledged), and *p* is the packet loss ratio which is the number of retransmitted packets divided by the total number of packets transmitted. Based on this formula the mean TCP Window size is *W*. With the formulation of this model, [30] proposes a similar equation to model WAN-related multi-stream TCP goodput and the authors investigate the verification of their formula. Based on the results in [30], then we can model and predict multi-stream TCP bandwidth with the relationship 2 in Table 1. In this equation, $n$ is the number of parallel TCP streams for an end-to-end transfer session and $BW_{agg}$ is the aggregate TCP throughput.

TABLE 2
WAN FORMULAS

| | |
|---|---|
| $BW = \dfrac{MSS}{RTT} \cdot \dfrac{C}{\sqrt{p}}$ | (1) |
| $BW_{agg} \leq C \cdot \sum_{i=1}^{n} \left( \dfrac{MSS_i}{RTT_i} \cdot \dfrac{1}{\sqrt{p_i}} \right)$ | (2) |
| $BW_{agg} \leq \dfrac{MSS}{RTT} \cdot n \cdot \dfrac{C}{\sqrt{p}} = k.n.W$ | (3) |
| $z = bandwidth = f(x, y) = f(n, W)$ | (4) |

In this table, $C/\sqrt{p}$ is equal to $W$ and $MSS/RTT$ is equal to $k$.

Using the approach and results of [30], if we assume that all the values of $RTT_i$, $P_i$ and $MSS_i$ are fixed, then for $n$ parallel streams, we obtain the equation 3 of Table 1. In relationship 3, $W$ is the representation of the TCP Window size. As seen from this relationship, the aggregate throughput rises with the increase in the values of $n$ and $W$.

This equation is valid for a specified interval. Increase or decrease in one of the $n$ or $W$ value affects the extensive variations in another parameter, which have been studied by many researchers to achieve a high throughput transfer rate. One of the chief reasons on this phenomenon may be the direct relation between TCP Window size and the packet loss formulated in equation 3 and the empirical relationship between packet loss and $n$ value.

For example, the excessive increase in the *n* parameter can be conducted to the TCP congestion phenomenon. This side effect increases packet loss due to dropping the TCP data segments exposed by direct interferences of gateways and routers in a typical network. Packet loss, however, may be due to other factors, such as intermittent hardware faults or physical layer distortions. In this section, we briefly express our experiences, related to measuring the throughputs of current DotDFS protocol implementation and Iperf tool, to validate the inferences of this model, to propose a new general abstraction to analytical model file transport protocols, and to optimally select the values of *n* and *W* in a typical WAN network with the testbed topology stated in section 5 (the main goal of this section is to model the behavior of DotDFS protocol and Iperf tool).

Since the most important effective parameters to achieve a high throughput rate in TCP-based WAN networks are *n* and *W* formulated in equation 3 of Table 1, we assume that the throughput is generally a function of two variables, *n* and *W* values (the equation 3 of Table 1 theoretically shows this fact). Therefore, we derive the equation 4 of Table 1.

As illustrated in equation 4, the main goal is to find a compact geometrical equation of a curve or surface which relates the variations of *z* coordinate according to the variations of *x* and *y* coordinates. When the implicit equation representing the *f* function can be found, the exterma or saddle points of the *f* function of two variables can be obtained, and therefore, we can find a set of points that maximize the *z* value corresponding to the variations of *x* and *y* coordinates. Ultimately, these obtained values are to represent the optimum values of *n* and *W*, which will contribute to a high throughput transfer rate for being used in Grid-based file transport protocols.

Fig. 20 and 21 show the three dimensional diagrams of the measured throughputs for DotDFS protocol and Iperf tool stated earlier in this section.

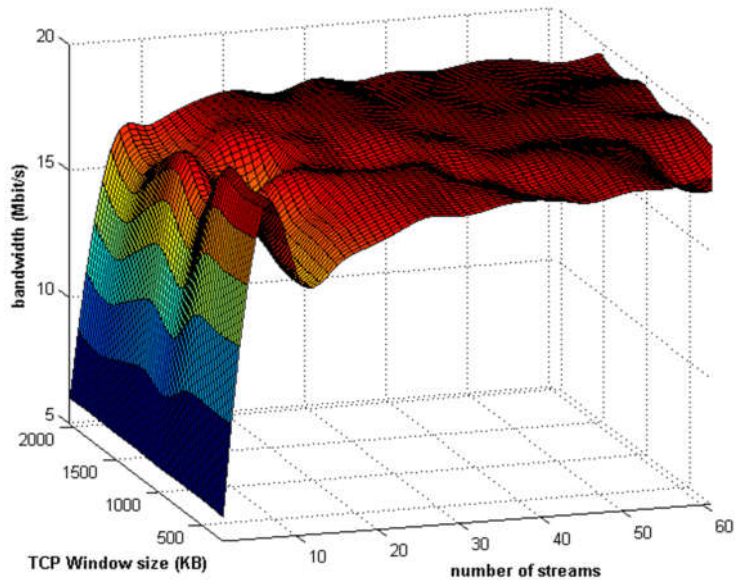

**Fig. 20.** 3D interpolated WAN Iperf throughput

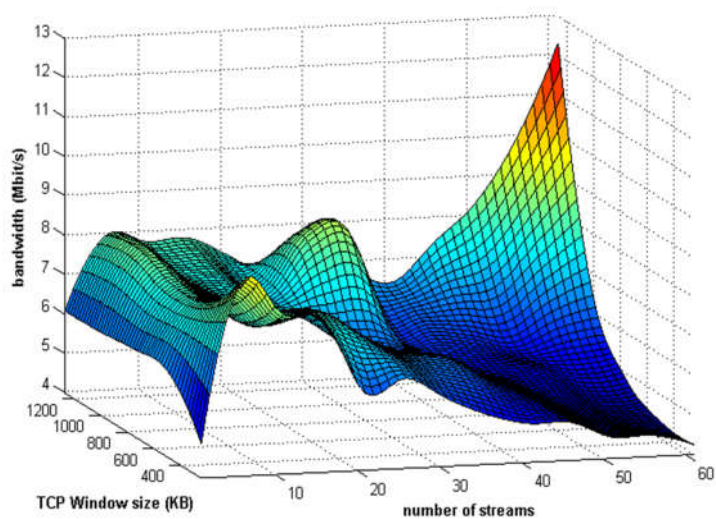

**Fig. 21.** 3D interpolated WAN DotDFS disk-to-disk throughput for a 1GB file transfer.

Diagrams 20 and 21 were obtained from nearly 500 collected raw data sets inspired by sampling, experimentation and mathematical interpolation approaches. Examining these diagrams reveals interesting results. From the most key inferences of this model are followed. Fig. 17 and 18 illustrate which the equation $z=f(x,y)$ is expressing an open surface with a set of exterma points. This set presents the points that maximize $z$ value along $x$ and $y$ variables meaning the pairs $(x,y)=(n,W)$. These pairs are the same points to achieve a high throughput transfer rate in our TCP-based WAN model. Equally, from these figures we can observe the validity of equations 1 to 3 for the total throughput of single and multiple (parallel) TCP streams. For example, the Fig. 20 shows to achieve a high WAN goodput and using the optimum network utilization the number of TCP streams must be set to 45 to 55 for a large TCP Window size such as 2MB for Iperf. Fig. 20 and 21 implicitly verify the same approaches proposed by Globus for accessing an optimal WAN goodput [31].

The final goal of key subjects covered in this section is to propose a new general model to optimal select $n$ and $W$ values and to develop a tool enabling the Grid developers to automatically measure these values without worrying about the network dynamic and the used Grid-based file transfer protocol like DotDFS or GridFTP soon.

## 8. Conclusion and Future Works

In this paper, we introduced the DotDFS protocol and expressed the characteristics that one can achieve to a Grid-based high-throughput file transfer system by using the DFSM, FTSM and PathM modes in the proposed and implemented protocol. We believe to achieve this goal, the procedures stated in this paper must be considered. Equally, we pointed out some weaknesses in the nature of the proposed GridFTP protocol and its implementation by the Globus team. The performance of the both .NET-based DotDFS and native Globus GridFTP were compared in different scenarios to validate the well-designed approaches used in the proposal and implementation of DotDFS protocol. Our LAN experiences in memory-to-memory tests showed that DotDFS accessed to the 94% bottleneck bandwidth while GridFTP was accessing 91%. In LAN disk-to-disk tests, comparison between DotDFS and GridFTP protocol unveiled a set of interesting and technical problems in GridFTP for both the nature of that protocol and its implementation by Globus.

We plan to do the following prioritized research works upon the DotDFS protocol based on results and experiences obtained and stated in this paper for the very near future:

1. The main future plan is to port the current .NET-based DotDFS implementation into the native runtime by using the standard C++ language in a cross-platform manner to be run-able in all operating systems. It will integrate the Win32/UNIX-compliant-POSIX APIs and utilize the prominent modern hardware advancements over the networks and storage systems. To probably prevent the objected-oriented overheads of C++ stack and to achieve the best optimized throughput, some critical cores of the new implemented DotDFS protocol will be written in pure C language.
2. Because of the verified overheads exposed by the widely-used TCP protocol, particularly in WAN networks upon file transport protocols, the other future work will be to provide a SCTP [32] stack driver for the underlying DotGridSocket interface to intercept SCTP packets instead of the TCP packets.
3. Another plan will be to expand the ideas stated in section 5.4 to analytical model the DotDFS protocol. We expect to provide an API-based network tool to optimally select the best network parameters such as the number of parallel streams and TCP Window buffer size in TCP-based transfer channels in pursuit of a high-throughput file transfer system.
4. And final, as stated during this paper, due to our knowledge we discovered an interesting phenomenon called the saturation speed. We plan to investigate in more detail what factors affect on throughput of large file transfers, especially for parallel streams and to find a solution for fixing the main problem.

We expect to provide and popularize a universal DotDFS framework for a variety of runtimes including .NET, Java, and C/C++ native code, which will unify the most available computing systems (desktop, server, notebook, netbook and mobile platforms) for highly secure, high-performance and high-throughput file transfers utilizing our protocol.

## References

[1] I. Foster, C. Kesselman, and S. Tuecke, The Anatomy of the Grid Enabling Scalable Virtual Organizations, International J. Supercomputer Applications, 15(3), 2001.
[2] I. Foster, C. Kesselman, J. Nick, and S. Tuecke, The Physiology of the Grid: An Open Grid Services Architecture for Distributed Systems Integration, Open Grid Service Infrastructure WG, Global Grid Forum, June 22, 2002.
[3] I. Foster, Globus Toolkit Version 4: Software for Service-Oriented Systems, IFIP International Conference on Network and Parallel Computing, Springer-Verlag LNCS 3779, pp 2-13, 2005.
[4] W. Allcock, J. Bresnahan, R. Kettimuthu, M. Link, C. Dumitrescu, I. Raicu and I. Foster, The Globus Striped GridFTP Framework and Server, in: Proceedings of Super Computing 2005 (SC05), November 2005.
[5] W. Allcock, GridFTP: Protocol extensions to ftp for the Grid, http://www.ggf.org/documents/GFD.20.pdf, 2010.
[6] B. Allcock, I. Mandrichenko, and T. Perelmutov, GridFTP v2 protocol description, http://www.ggf.org/documents/GFD.47.pdf, 2010.
[7] J. Feng, L. Cui, G. Wasson, and M. Humphrey, Toward Seamless Grid Data Access: Design and Implementation of GridFTP on .NET, in: Proceedings of the 2005 Grid Workshop, Seattle, USA, Nov 13-14, 2005.
[8] R. Kalmady, and B. Tierney, A Comparison of GSIFTP and RFIO on a WAN, in: Proceedings of CHEP'01, Beijing, China, Sept 3-7, 2001.
[9] T. Baer, and P. Wyckoff, A parallel I/O mechanism for distributed systems, in: Proceedings of the 2004 IEEE International Conference on Cluster Computing, California, USA, Sept 20-23, 2004.
[10] R. Niederberger, W. Allcock, L. Gommans, E. Grünter, T. Metsch, I. Monga, G. L. Volpato, and C. Grimm, Firewall Issues Overview, http://www.ggf.org/documents/GFD.83.pdf, 2010.
[11] T. Metsch, L. Gommans, E. Grünter, R. Niederberger, A. de Smet, and G. L. Volpato, Requirements on operating Grids in Firewalled Environments, http://www.ggf.org/documents/GFD.142.pdf, 2010.
[12] A. Poshtkohi, A.H. Abutalebi, and S. Hessabi, DotGrid: a .NET-based cross-platform software for desktop grids, International Journal of Web and Grid Services 2007, Vol. 3, No.3 pp. 313 – 332.
[13] A. Poshtkuhi, A. Abutalebi, L. Ayough, and S. Hessabi, DotGrid: A .NET-based Infrastructure for Global Grid Computing, in: Proceedings of the 6th IEEE International Symposium on Cluster Computing and the Grid (CCGrid2006), 16-19 May 2006, Singapore.
[14] A. Poshtkuhi, A. Abutalebi, L. Ayough, and S. Hessabi, DotGrid: A .NET-based Cross-Platform Grid Computing Infrastructure, in: Proceedings of the IEEE International Conference On Computing and Informatics 2006 (ICOCI06), Malaysia, June 6-8, 2006.

[15] Microsoft Corporation, .NET Framework Home, http://msdn.microsoft.com/netframework/, 2010.
[16] MONO .NET Project Home Page, http://www.mono-project.com/, 2010.
[17] Z. Peter Kunszt and P. Leanne Guy, The Open Grid Services Architecture, and data Grids, Grid Computing Making the Global Infrastructure a Reality, John Wiley, 2003.
[18] Lustre Cluster File System, http://www.lustre.org/, 2010.
[19] General Parallel File System (GPFS), http://www-1.ibm. com/servers/eserver/clusters/software/gpfs.html, 2010.
[20] Gfarm File System, http://datafarm.apgrid.org/, 2010.
[21] V. Welch, F. Siebenlist, I. Foster, J. Bresnahan, K. Czajkowski, J. Gawor, C. Kesselman, S. Meder, L. Pearlman, and S. Tuecke, Security for Grid Services, Twelfth International Symposium on High Performance Distributed Computing, IEEE Press, June 2003.
[22] I. Foster, C. Kesselman, G. Tsudik, and S. Tuecke, A security architecture for computational Grids, in: Proceedings of the 5th ACM Computer and Communications Security Conference (CCS '98), pages 83–92. ACM Press, November 1998.
[23] T. Dierks, and C. Allen, The TLS protocol version 1.0, IETF, 1999.
[24] J. Crampton, H. W. Lim, K. G. Paterson, and G. Price, A certificate-free grid security infrastructure supporting password-based user authentication, in: Proceedings of the 6th Annual PKI R&D Workshop 2007, NIST, 2007.
[25] J. Clark, Calling Win32 DLLs in C# with P/Invoke, MSDN Magazine, The Microsoft Journal for Developers, July, 2003, http://msdn.microsoft.com/msdnmag/issues/03/07/NET/default.aspx, 2010.
[26] Bonnie file system benchmark, http://www.textuality.com/bonnie, 2010.
[27] V.S. Pai, P. Druschel, and W. Zwaenepoel, Flash: An efficient and portable web server, in: Proceedings of the USENIX 1999 Annual Technical Conference, June 1999.
[28] J. Bresnahan, M. Link, R. Kettimuthu, D. Fraser and I. Foster, GridFTP Pipelining, in: Proceedings of the 2007 TeraGrid Conference, June, 2007.
[29] M. Mathis, J. Semke, J. Mahdavi, and T. Ott, The Macroscopic Behavior of the TCP Congestion Avoidance Algorithm, Computer Communication Review, volume 27, number3, July 1997.
[30] T.J. Hacker, B.D. Athey, and B. Noble, The End-to-End Performance Effects of Parallel TCP Sockets on a Lossy Wide-Area Network, in: Proceedings of the 16th International Parallel and Distributed Processing Symposium, Florida, USA, April 15-19, 2004.
[31] J. Lee, D. Gunter, B. Tierney, B, Allcock, J. Bester, J. Bresnahan, S. Tuecke, Applied Techniques for High Bandwidth Data Transfers Across Wide Area Networks, in: Proceedings of International Conference on Computing in High Energy and Nuclear Physics, Beijing, China, September 2001.
[32] L. Ong, and J. Yoakum, An Introduction to the Stream Control Transmission Protocol (SCTP), IETF, RFC 3286, 2002.
[33] ECMA-334: C# Language Specification, http://www.ecma-international.org/publications/standards/Ecma-334.htm, 2010.
[34] ECMA-335: Common Language Infrastructure (CLI), http://www.ecma-international.org/publications/techreports/E-TR-084.htm, 2010.